\documentclass[12pt]{spieman}  
\usepackage{amsmath,amsfonts,amssymb}
\usepackage{graphicx}
\usepackage{setspace}
\usepackage{tocloft}
\usepackage{enumitem}

\title{Development of the Timing System for the X-Ray Imaging and Spectroscopy Mission}

\author[a,b,*]{Yukikatsu Terada}
\author[c]{Megumi Shidatsu}
\author[d,e]{Makoto Sawada}
\author[f]{Takashi Kominato}
\author[a]{So Kato}
\author[a]{Ryohei Sato}
\author[a]{Minami Sakama}
\author[a]{Takumi Shioiri}
\author[c]{Yuki Niida}
\author[b]{Chikara Natsukari}
\author[a,b]{Makoto S. Tashiro}
\author[b]{Kenichi Toda}
\author[b]{Hironori Maejima}
\author[b]{Katsuhiro Hayashi}
\author[b]{Tessei Yoshida}
\author[b]{Shoji Ogawa}
\author[b]{Yoshiaki Kanemaru}
\author[b]{Akio Hoshino}
\author[b]{Kotaro Fukushima}
\author[g]{Hiromitsu Takahashi}
\author[h]{Masayoshi Nobukawa}
\author[g]{Tsunefumi Mizuno}
\author[i]{Kazuhiro Nakazawa}
\author[j]{Shin'ichiro Uno}
\author[b]{Ken Ebisawa}
\author[k]{Satoshi Eguchi}
\author[a]{Satoru Katsuda}
\author[l]{Aya Kubota}
\author[m]{Naomi Ota}
\author[n]{Atsushi Tanimoto}
\author[c]{Yuichi Terashima}
\author[o]{Yohko Tsuboi}
\author[p]{Yuusuke Uchida}
\author[q]{Hideki Uchiyama}
\author[m]{Shigeo Yamauchi}
\author[o]{Tomokage Yoneyama}
\author[r]{Satoshi Yamada}
\author[b]{Nagomi Uchida}
\author[b]{Shin Watanabe}
\author[b]{Ryo Iizuka}
\author[b]{Rie Sato}
\author[s]{Chris Baluta}
\author[s]{Matt Holland}
\author[t,s]{Michael Loewenstein}
\author[u]{Eric D. Miller}
\author[v,s]{Tahir Yaqoob}
\author[w]{Robert S. Hill}
\author[w]{Morgan D. Waddy}
\author[w]{Mark Mekosh}
\author[w]{Joseph B. Fox}
\author[w]{Emily Aldoretta}
\author[w]{Isabella Brewer}
\author[v,s]{Koji Mukai}
\author[v,s]{Kenji Hamaguchi}
\author[t,s]{Francois Mernier}
\author[t,s]{Anna Ogorzalek}
\author[v,s]{Katja Pottschmidt}
\author[x,s]{Mihoko Yukita}
\author[c]{Toshihiro Takagi}
\author[a]{Yugo Motogami}  
\author[y]{Teruaki Enoto}
\author[z]{Takaaki Tanaka}
\author[c]{Taichi Nakamoto}
\author[c]{Chulsoo Kang}
\author[a]{Tsuyoshi Miyazaki}  

\affil[a]{Saitama University, Graduate School of Science and Engineering, 255 Shimo-Ohkubo, Sakura-ku, Saitama-shi, Saitama, 338-8570, Japan} 
\affil[b]{Japan Aerospace Exploration Agency, Institute of Space and Astronautical Science, 3-1-1, Yoshinodai, Chuo-ku, Sagamihara, Kanagawa, 252-5210, Japan} 
\affil[c]{Ehime University, Graduate School of Science and Engineering, 2-5, Bunkyo-cho, Matsuyama, Ehime, 790-8577, Japan} 
\affil[d]{Rikkyo University, Department of Physics, 3-34-1, Nishi-Ikebukuro, Toshima-ku, Tokyo, 171-8501, Japan}
\affil[e]{RIKEN, Center for Pioneering Research, 2-1 Hirosawa, Wako, Saitama 351-0198, Japan} 
\affil[f]{NEC Corp., 1-10-2, Nishin-tyou, Fuchu-shi, Tokyo, 183-0036, Japan}
\affil[g]{Hiroshima University, School of Science, 1-3-2, Kagamiyama, Higashi-Hiroshima, Hiroshima, 739-0046, Japan}
\affil[h]{Nara University of Education, Faculty of Education, Takabatake-cho, Nara-shi, Nara, 630-8301, Japan} 
\affil[i]{Nagoya University, Department of Physics, Furo-cho, Chikusa-ku, Nagoya, Aichi, 464-8601, Japan}  
\affil[j]{Nihon Fukushi University, Faculty of Health Sciences, 26-2, Higashi-Ikumi, Handa, Aichi, 475-0012, Japan}
\affil[k]{Kumamoto Gakuen University, Faculty of Economics, Department of Economics, 2-5-1 Oe, Chuo-ku, Kumamoto, Japan, 862-8680}  
\affil[l]{Shibaura Institute of Technology, Department of Electronic Information Systems, 307 Fukasaku, Minuma-ku, Saitama-shi, Saitama, 337-8570, Japan}
\affil[m]{Nara Women’s University, Department of Physics, Kitauoya-nishi, Nara-shi, Nara, 630-8506, Japan} 
\affil[n]{Kagoshima University, Faculty of Science, 1-21-24, Kohrimoto, Kagoshima-shi, Kagoshima, 890-0065, Japan}  
\affil[o]{Chuo University, Department of Physics, Bunkyo, Tokyo, Japan}
\affil[p]{Tokyo University of Science, 2641 Yamazaki, Noda-shi, Chiba, 278-8510, Japan}  
\affil[q]{Shizuoka University, Faculty of Education, 836, Ohya, Suruga-ku, Shizuoka-shi, Shizuoka, 422-8529, Japan} 
\affil[r]{RIKEN, Nishina Center, 2-1 Hirosawa, Wako, Saitama, 351-0198, Japan}
\affil[s]{National Aeronautics and Space Administration (NASA), Goddard Space Flight Center, 8800 Greenbelt Road, Greenbelt, Maryland, 20771, United States}
\affil[t]{University of Maryland, College Park, Maryland, 20742, United States}
\affil[u]{Massachusetts Institute of Technology, Kavli Institute for Astrophysics and Space Research, 77 Massachusetts Avenue, Cambridge, Massachusetts, 02139, United States} 
\affil[v]{University of Maryland, Baltimore County, 1000 Hilltop Circle, Baltimore, Maryland, 21250, United States}  
\affil[w]{ADNET Systems Inc., 6720B Rockledge Drive, Suite \# 504. Bethesda, Maryland, 20817, United States} 
\affil[x]{Johns Hopkins University, 3400 N. Charles Street Baltimore, Maryland, 21218, United States}
\affil[y]{Kyoto University, Yoshida-honmachi, Sakyo-ku, Kyoto 606-8501, Japan}
\affil[z]{Konan University, 8-9-1 Okamoto, Higashinada, Kobe, Hyogo, 658-8501, Japan}

\cftpagenumbersoff{figure}
\cftpagenumbersoff{table} 

\newcommand{\etal}{{\it et al.\ }}
\newcommand{\aap}{\it Astronomy \& Astrophysics}
\newcommand{\pasj}{\it Publications of the Astronomical Society of Japan}
\newcommand{\apj}{\it Astrophysical Journal}
\newcommand{\apjl}{\it Astrophysical Journal Letter}

\begin{document} 
\maketitle

\begin{abstract}
This paper describes the development, design, ground verification, and in-orbit verification, performance measurement, and calibration of the timing system for the X-Ray Imaging and Spectroscopy Mission (XRISM).
The scientific goals of the mission require an absolute timing accuracy of 1.0~ms.
All components of the timing system were designed and verified to be within the timing error budgets, which were assigned by component to meet the requirements.
After the launch of XRISM, the timing capability of the ground-tuned timing system was verified using the millisecond pulsar PSR~B1937+21 during the commissioning period, and the timing jitter of the bus and the ground component were found to be below $15~\mu$s compared to the NICER (Neutron star Interior Composition ExploreR) profile.
During the performance verification and calibration period, simultaneous observations of the Crab pulsar by XRISM, NuSTAR (Nuclear Spectroscopic Telescope Array), and NICER were made to measure the absolute timing offset of the system, showing that the arrival time of the main pulse with XRISM was aligned with that of NICER and NuSTAR to within $200~\mu$s.
In conclusion, the absolute timing accuracy of the bus and the ground component of the XRISM timing system meets the timing error budget of $500~\mu$s.
\end{abstract}

{\noindent \footnotesize\textbf{*}Yukikatsu Terada,  \linkable{terada@mail.saitama-u.ac.jp} }

\begin{spacing}{1}   %


\section{Introduction: Timing requirement for XRISM}
\label{sec:timing_req}  

The X-Ray Imaging and Spectroscopy Mission (XRISM) is an X-ray astrophysical satellite mission led by the Japan Aerospace Exploration Agency (JAXA) and the National Aeronautics and Space Administration (NASA) in collaboration with the European Space Agency (ESA) and other international partners.\cite{2020SPIE11444E..22T, 2024SPIE130931G_XRISM}
The scientific objectives of the mission are to understand the structure and evolution of the universe, the circulation history of baryonic matter, and the transport of energies by making high-resolution X-ray observations of astronomical objects, and to recover high-resolution science after the loss of the Hitomi mission.\cite{2018JATIS...4b1402T}
The spacecraft carries an X-ray microcalorimeter array (known as Resolve)\cite{2022SPIE12181E..1SI,2024SPIE_Kelley} and an X-ray CCD camera (known as Xtend)\cite{2022SPIE12181E..1TM,2024SPIE_Mori} in the focal planes of its X-ray mirrors. 
These were designed to perform X-ray spectroscopy with a high energy resolution of $\le7$~eV FWHM in a field of view (FOV) of $3.05 \times 3.05$ arcmin$^2$ and X-ray imaging capability in a wide FOV of $38 \times 38$ arcmin$^2$ in the 0.3--13~keV and 0.4--12~keV bands, respectively.
In the initial year after launch, the gate valve (GV) protecting the Resolve sensor with its thin beryllium entrance window remains closed, which limits the energy bandpass of Resolve to 1.7--12~keV.
The mission serves as a public observatory, providing global access for research scientists.
To support these guest observers, the science operations for the mission---e.g., science observation planning, on-ground pre-pipeline and pipeline processes, archiving, user support---were designed and prepared before launch\cite{2021JATIS...7c7001T} and started soon thereafter.\cite{2024SPIE_Hayashi,2025JATIS_Hayashi_SOC}

The mission requirements were defined by the scientific objectives of the mission.
XRISM will address a wide range of scientific objectives, together with its main science goals, for which not only spectral and imaging functions but also timing capabilities are very important.
The requirement on the absolute timing accuracy for XRISM is 1.0~ms with $1\sigma$ deviation.
This value is a relaxation from that for the Hitomi timing system as summarized in Table~\ref{tab:time_requirement}.
The mission requirements include the performance of both the payload instruments and the common components, which include the bus and ground component.
For the timing performance, the requirement is focused on the instruments with the highest precision; for Hitomi, those were the high-energy instruments (HXI and SGD),\cite{2018JATIS...4a1206T} whereas for XRISM it is the microcalorimeter array (Resolve), prompting an update in mission requirements.
Separately, based on the timing calibration result for Hitomi SXS,\cite{2018JATIS...4b1407L} the subsystem requirement for the microcalorimeter array (10~ms)\cite{2018JATIS...4b1406E} has also been changed and is now tighter.  

\begin{table}[hbt]
    \centering
    \caption{Timing requirements for subsystems on Hitomi and XRISM satellites.}
    \begin{tabular}{ccccl}
    \hline 
     & \multicolumn{2}{c}{\bf Hitomi} & \multicolumn{2}{c}{\bf XRISM} \\
    \hline 
    \multicolumn{4}{l}{\bf Mission requirement}\\
                                &\multicolumn{2}{c}{350~$\mu$s absolute\cite{2018JATIS...4a1206T}} 
                                &\multicolumn{2}{c}{1.0~ms absolute$^\dagger$} \\
    \hline 
    \multicolumn{4}{l}{\bf Subsystem requirements}\\
    Microcalorimeter   & SXS & 10~ms absolute\cite{2018JATIS...4b1406E} & Resolve   & 500~$\mu$s absolute$^\dagger$\\
    X-ray CCD camera    & SXI & 61~$\mu$s relative$^\ddagger$       & Xtend     & 10~ms absolute\\
    Hard X-ray Instrument & HXI & 61~$\mu$s relative$^\ddagger$     &\multicolumn{2}{c}{N/A} \\
    Soft Gamma-ray Instrument & SGD & 61~$\mu$s relative$^\ddagger$ &\multicolumn{2}{c}{N/A} \\
    Timing system      & Bus+Ground  & 350~$\mu$s absolute           & Bus+Ground   &  500~$\mu$s absolute$^\dagger$\\
    \hline 
    \end{tabular}
    \\
    {\scriptsize $\dagger$:  Half is assigned for SXS (see the text).}
    {\scriptsize $\ddagger$: Timing accuracy of payload instruments relative to bus part of timing system.}
    \label{tab:time_requirement}
\end{table}

For the XRISM timing system, this paper describes its development, design, ground verification, and in-orbit verification and performance measurement, following on from initial results reported by the SPIE proceeding, Terada \etal (2024)\cite{2024SPIE_Terada}.
As noted above, the mission requirement is based on Resolve, so omitted herein are the verification and performance measurement of Xtend; see future work for the specifics of its timing performance.
The rest of this paper is as follows. Section~\ref{sec:timing_system} describes the on-ground design and development of the XRISM timing system. Sections~\ref{sec:ground_test} and \ref{sec:timing_commissioning} report the verification of the timing system both on the ground and in orbit, respectively. Section~\ref{sec:timing_crab} reports the measurement of the in-orbit performance.

\section{Design and Development of Timing System}
\label{sec:timing_system}  
\subsection{Design}
\label{sec:timing_system_design}  
\begin{figure}[hbt]
    \centering
   \includegraphics[width=0.80 \textwidth]{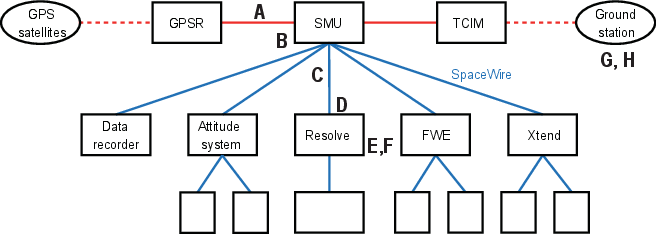}
    \caption{Schematic of the logical topology of the XRISM network, updated from the Hitomi case in Fig.~1 of Terada \etal (2018)\cite{2018JATIS...4a1206T}.
    Ellipses represent GPS satellites or the ground station, and boxes are components onboard the spacecraft, where the empty boxes are sub-system components within a local network of Attitude system, Resolve, FWE, or Xtend systems.
    Communication lines (in blue) are realized by SpaceWire.
    Also shown are the timing error IDs from Table~\ref{tab:error_budget}.}
    \label{fig:time_system_xrism}
\end{figure}

XRISM uses the same timing system as Hitomi, so the system already has a higher absolute timing accuracy of $350~\mu$s\cite{2018JATIS...4a1206T} compared with the XRISM science requirement of 1.0~ms absolute timing accuracy as described in Sec.~\ref{sec:timing_req}.
As shown in Fig.~\ref{fig:time_system_xrism}, the XRISM timing system involves the spacecraft and a ground system.
The spacecraft carries a GPS receiver (GPSR) that provides accurate timing to a few tens of nanoseconds in the International Atomic Time (TAI) system to the onboard central computer known as the Satellite Management Unit (SMU).
The spacecraft time is known as the time indicator (TI) and is distributed from the SMU to the payload instruments via the SpaceWire network (IEEE1355)\cite{SpaceWire:IEEE1355,SpaceWire:D} through telemetry and command communications.
From the payload instruments (i.e., SpaceWire nodes in Fig.~\ref{fig:time_system_xrism}), the TI information of space packets is sent to the ground system via SMU, and the final timing information of the space packets or X-ray events---known as TIME---is calculated in the pre-pipeline and pipeline processes on the ground\cite{2018JATIS...4a1206T,2021JATIS...7c7001T} using the LOCAL\_TIME and the TI information\cite{AHTIME} in the telemetry, and stored into the output files in the flexible image transport system (FITS) format\cite{2001A&A...376..359H}, as described in detail by Terada \etal (2018)\cite{2018JATIS...4a1206T}

\subsection{Timing Error Budget Control in Development Phase}
\label{sec:timing_system_budget}  

\begin{table}[htb]
    \centering
    \caption{Error budget for timing accuracy.}
    \begin{tabular}{ccccc}
    \hline 
    ID  & Component/Line          & Hitomi budget & Hitomi performance & XRISM budget \\
    \hline 
    {\bf A}   & GPSR - SMU    &   $0.02~\mu$s & $0.01~\mu$s   & $0.2~\mu$s \\
    {\bf B}   & SMU - SpW network & $0.5~\mu$s & $0.14~\mu$s  & $350~\mu$s \\
    {\bf B}$^\dagger$   & SMU - SpW network & $270~\mu$s & N/A & $350~\mu$s \\
    {\bf C}   & SpW network &     $2.0~\mu$s  & $0.1$--$1.0~\mu$s & $2.0~\mu$s \\
    {\bf D}   & SpW network - Instrument & $1.0~\mu$s & $<1.0~\mu$s   & $1.0~\mu$s $^\ddagger$\\
    {\bf E}   & Instrument    &   $1.0~\mu$s  & $<0.6~\mu$s   & $2.0~\mu$s $^\ddagger$ \\
    {\bf F}   & LOCAL\_TIME   &   $5$--$25.6~\mu$s & $5$--$25.6~\mu$s   & $132~\mu$s (H/M)$^\ddagger$\\
        &               &                      &                     & $207~\mu$s (Lp) $^\ddagger$ \\
        &               &                      &                     & $242~\mu$s (Ls) $^\ddagger$ \\
    {\bf G}   & Orbital Element & $3.0~\mu$s   & $0.0005~\mu$s & $3.0~\mu$s \\
    {\bf H}   & Systematic in time ref.\ &    N/A & N/A&  $ 150~\mu$s $^\ddagger$\\
    \hline 
    \end{tabular}
    \\
    {\scriptsize $\dagger$: GPSR failure case in GPS~OFF mode in Terada \etal (2018)\cite{2018JATIS...4a1206T}.} {\scriptsize $\ddagger$: error budget for Resolve instrument\cite{2025JATIS_Sawada_Time}.}
    \label{tab:error_budget}
\end{table}

During the propagation of timing information from TAI to TIME, the timing jitters and/or uncertainties degrade the total timing accuracy of TIME.
In the development of the Hitomi timing system, seven error items were identified that affect the timing accuracy (see Fig.~2 of Terada \etal 2018\cite{2018JATIS...4a1206T}).
The error items and budgets for the XRISM timing system are summarized in Table~\ref{tab:error_budget}.
Because Hitomi's requirement on absolute timing accuracy has been relaxed in XRISM, the budgets for items A, B, E, and F have been updated in XRISM.
Also, we explicitly defined the new item~H, which is the systematic uncertainty about the timing offset for Resolve applied to the end-to-end timing performance.
Resolve events are categorized by grades based on the time interval $\delta$ between events\cite{2022SPIE12181E..1SI}. 
The budget for item F is determined by such event grades. 
Grades H, M, and L are assigned as follows: if $\delta >$ 70.72 ms, then H resolution; if 18.32 ms $< \delta \leq $ 70.72 ms, then M resolution; and if $\delta \leq$ 18.32 ms, then L resolution. 
Leading and trailing events receive primary and secondary flags, respectively. 
These grades and flags combine to form the five event types: Hp, Mp, Ms, Lp, and Ls.

During the on-ground development and verification, these error budgets were controlled by two subsystems: one for the Resolve instrument ($500~\mu$s) and the other for the bus and ground component ($500~\mu$s) of the timing system.
In more detail, items A, B, C, and G are assigned to the bus and ground component, and items D, E, F, and G are assigned to the Resolve instrument. 
The detailed implementation for the Resolve items is discussed in Sawada \etal (2025)\cite{2025JATIS_Sawada_Time}, in which, items E and F are split into two and three sub-items, respectively: E-1, which addresses the uncertainty in total propagation delay of TI within the SpaceWire network, and E-2, focusing on jitter within the user node of Resolve, F-1 concerns trigger time resolution, F-2 addresses synchronization errors between Resolve components, and F-3 deals with systematic uncertainty in absolute timing reference as a margin.
Herein, we describe the bus and ground component of the XRISM timing system.

\subsection{Development Steps of Timing System}
\label{sec:timing_system_develop}  

For a spacecraft and its ground system, development (i.e., design, fabrication, and verification) is generally completed before launch, after which function checks under full flight configuration in orbit are performed in the commissioning period of the initial operation phase after the critical operation period.
The development of the present timing system followed this normal procedure of systems engineering.
In detail, the verification and calibration of the timing system were performed in the following three steps.
Here, note that the performance verification (PV) and calibration period and the guest observation period follow the commissioning phase after the launch of XRISM, as the normal operation phase.\cite{2021JATIS...7c7001T}

\begin{enumerate}[label=\textbf{Step \arabic*}]
    \item On the ground before launch, we designed and fabricated the timing system so that each component would satisfy the timing error budget listed in Table~\ref{tab:error_budget}, and we verified that the timing performance of each component was within budget.
    \item In the commissioning phase soon after launch, the overall (end-to-end) timing performance verification using a millisecond pulsar were performed to check no degradation from the ground expectation.
    \item In the PV and calibration period during nominal operations, along with other X-ray observatories, we performed simultaneous observations of the Crab Pulsar neutron star as a standard candle in order to measure/calibrate the absolute timing accuracy with the final timing parameters tuned on the ground.
\end{enumerate}
The GPS~OFF mode of item~B was not verified explicitly in the Hitomi mission, but we did so for XRISM before launch in step~1, as reported by Shidatsu \etal (2024, 2025)\cite{2024SPIE_Shidatsu,2025JATIS_Shidatsu_Time}.
The details of steps 1, 2, and 3 are described in Secs.~\ref{sec:ground_test}, \ref{sec:timing_commissioning}, and \ref{sec:timing_crab}, respectively.

\section{Timing Verification Experiment on Ground}
\label{sec:ground_test}  
\subsection{Purpose}
\label{sec:ground_test_purpose}  

We performed a timing verification experiment on the ground as in step~1 defined in Sec.~\ref{sec:timing_system_develop}.
The main purpose was to verify that the combined timing performance of items A, B, and C is within the error budget ($351.2~\mu$s; Table~\ref{tab:error_budget}).

\subsection{Configuration}
\label{sec:ground_test_config}  

Verification tests were performed in the following three configurations:
\begin{itemize}
    \item[{\bf (a)}] subsystem configuration of SMU with flight bus components at room temperature before delivery to JAXA;
    \item[{\bf (b)}] as (a) but with SMU and related components now on board the spacecraft and at room temperature;
    \item[{\bf (c)}] full flight configuration in thermal vacuum chamber under hot/cold conditions.
\end{itemize}
The tests in configurations (a), (b), and (c) were performed on 26--28 January 2021 at a test facility of the NEC Corporation in Tokyo, on 13 September 2021 at the JAXA Tsukuba Space Center (TKSC), and from 4 August to 1 September 2022 at JAXA TKSC, respectively.

\begin{figure}[hbt]
    \centering
   \includegraphics[width=0.85 \textwidth]{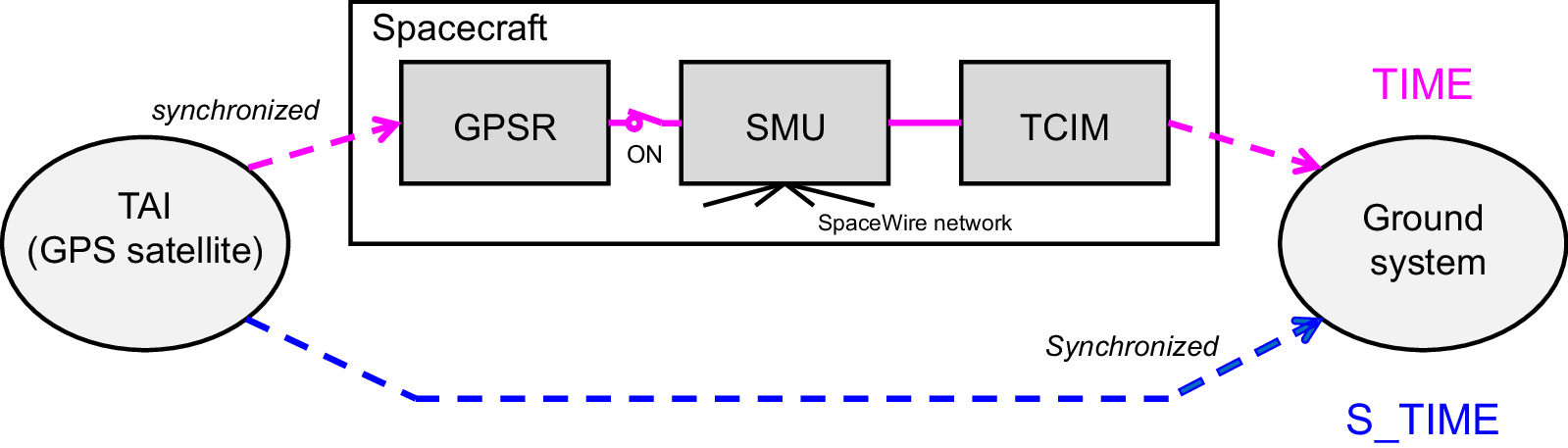}
    \caption{Schematic of the configuration of the ground verification experiment for timing.}
    \label{fig:ground_setup}
\end{figure}

The experimental configuration is shown schematically in Fig.~\ref{fig:ground_setup}.
The GPSR on board the spacecraft was set to ground operation mode (i.e., no in-orbit conditions and without Doppler correction from GPS signals), but the output clock from GPSR to SMU was synchronized with TAI. 
Therefore, the onboard TI was expected to be synchronized with TAI via GPSR.
As a target of this verification, TIME was calculated from TI and other telemetry information, as described in Sec.~\ref{sec:timing_system_design}.
In this verification experiment, the ground system was also synchronized with TAI, and therefore the packet receive time (S\_TIME) was synchronized with TAI and could be used as a reference time when examining TIME.

We tested both the GPS~ON and GPS~OFF modes defined in Terada \etal (2018)\cite{2018JATIS...4a1206T}. 
In GPS~ON mode, TI is synchronized with TAI by connecting the GPSR signal to SMU as indicated in Fig.~\ref{fig:ground_setup}, while in GPS~OFF mode, the TI information is generated by the free-running clock on SMU.
In real operation, GPS~OFF mode is realized automatically when GPSR fails to obtain signals from the GPS satellite or there are other failures in the delivery of the timing signal from GPSR to SMU.
Herein, we concentrate on reporting the results for GPS~ON mode, and Shidatsu \etal (2024, 2025)\cite{2024SPIE_Shidatsu,2025JATIS_Shidatsu_Time} describes the details of the setup and results of the verification test for GPS~OFF mode.

\subsection{Results}
\label{sec:ground_test_result}  
\begin{figure}[hbt]
    \centering
    \includegraphics[width=0.95 \textwidth]{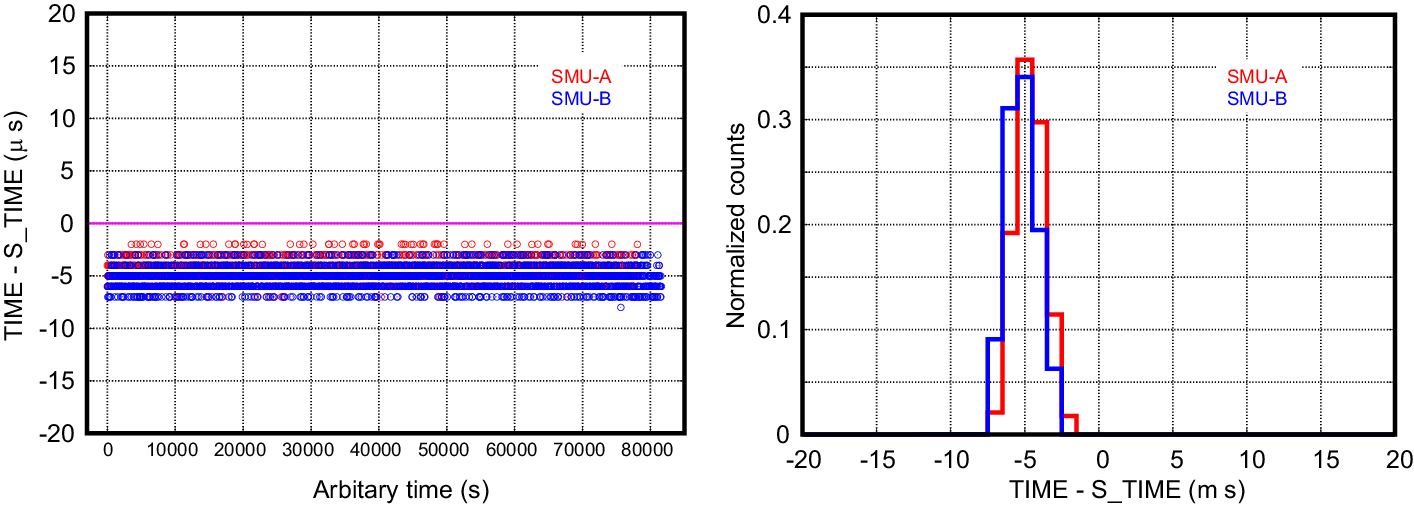}
    \caption{Left: trend of timing difference between TIME and S\_TIME, taken in ground verification test in configuration~(c) (see the text); red and blue represent the results of SMU-A and SMU-B, respectively. Right: distribution of difference between TIME and S\_TIME shown on left.}
    \label{fig:ground_result}
\end{figure}

First, we extracted the time packet from the telemetry generated by the onboard bus component (TCIM in Fig.~\ref{fig:ground_setup}), which is used for communication between the spacecraft and the ground station.
The latch timing of the TI counter in the time packet is well-defined with fixed timing delay and negligible timing jitters (a few microseconds).
This timing delay consists of the propagation delay from the spacecraft to the ground station, which is calculated from the orbital element, and the fixed delay on the ground equipment, which depends on the data link rate for the communication.
Note that TI values in normal space packets---such as housekeeping (HK) telemetries and science event telemetries---are not suitable for timing verification within a few hundred microseconds because they are attached by the onboard software of SMU in the editing process of the space packets and thus have large timing jitters and an uncontrolled delay on the order of 10~ms.
We then performed the time-assignment calculation for TI in the time packet to extract the TIME information as a result.

Figure~\ref{fig:ground_result} shows the results of the ground timing verification test in configuration~(c) (i.e., the final flight configuration in the spacecraft thermal vacuum experiment).
As can be seen, the temporal discrepancy between TIME and the reference time S\_TIME was maintained within $10~\mu$s over a duration of about 80~ks for each of the two redundant SMU systems.
The cause of the observed residual of about $-5~\mu$s in the figure remains unidentified.
Note that this unknown residual is not subtracted in the XRISM timing system, because it is negligible compared with the error budget for item B and it falls within the timing resolution of TI in the time packet.
Therefore, we conclude that the timing performance in nominal operation mode (i.e., GPS~ON) satisfies the error budget for items A, B, and C.

\section{In-orbit Timing Verification During Commissioning Period in Initial Operation Phase}
\label{sec:timing_commissioning}  

\subsection{Purpose}
\label{sec:timing_commissioning_purpose}  

We verified the timing in orbit during the commissioning period in the initial operation phase, as in step~2 defined in Sec.~\ref{sec:timing_system_develop}.
The main purpose of this timing verification in the commissioning period was to confirm that (i) there is no degradation in the bus components of the timing system and (ii) we can measure the absolute timing accuracy of Resolve high-resolution primary (Hp) or medium-resolution primary (Mp) grade events in the later nominal operation phase.
Note that confirming the former item is sufficient for GPS~OFF mode because this is a backup mode, and so there was no request to check the timing performance in GPS~OFF mode during the commissioning period\cite{2024SPIE_Shidatsu,2025JATIS_Shidatsu_Time}.
To confirm the nominal timing mode (GPS~ON), we planned fast-rotating pulsar observations as described below.

\subsection{Observation Summary}
\label{sec:timing_commissioning_observation}  

Before launch, we planned the details of the timing verification commissioning, listing neutron star pulsars as candidates for timing verification observation during the initial operation phase.
The objects were selected based on the pulse period $P$, the sharpness of the pulse profile, and the same photon statistics with an exposure of 80~ks with Resolve (planned for an open GV condition; Section \ref{sec:timing_req}).
The candidates are summarized in Table~\ref{tab:commissioning_objects}, ordered by priority.

\begin{table*}[hbt]
    \centering
    \caption{Candidates for timing verification commissioning.}
    \begin{tabular}{lccccl}
    \hline 
    Object name & RA & DEC              & $P^\dagger$ & Hp rate$^\ddagger$ & Visibility\\
    \hline 
    Crab        &83.633080 & 22.014500  & 33.7  & 30/42         & 17 Aug -- 18 Oct 2023,\\
    & & & & & 13 Feb -- 13 Apr 2024\\
    PSR~B1821-24&276.133371&-24.869750  & 3.05  & 0.020/0.006   & 29 Aug -- 29 Oct 2023,\\
    & & & & & 25 Feb -- 25 Apr 2024\\
    PSR J1937+21&294.915100&21.622700   & 1.56  & 0.011/0.004   & 12 Sep -- 7 Dec 2023,\\
    & & & & & 9 Mar -- 6 Jun 2024\\
    PSR J0218+4232 &34.526502 &42.538157 & 2.32 & 0.023/0.006   & 5 Jul -- 14 Sep 2023,\\
    & & & & & 4 Jan -- 11 May 2024\\
    \hline 
    \end{tabular}
    \\
    {\scriptsize $\dagger$ Pulse period in milliseconds. 
    $\ddagger$ Count rate expected for Resolve Hp grade in units of counts per second under gate-valve closed/open conditions.}
    \label{tab:commissioning_objects}
\end{table*}

Not all objects are observable with XRISM in one epoch because the spacecraft has a fixed solar paddle and its attitude is limited on a given date. 
Accounting for such visibility epochs for the candidates listed in Table~\ref{tab:commissioning_objects}, we observed PSR~B1937+21 (observation ID (OBSID) = 000123000) for timing verification commissioning with a net exposure of 245~ks on 25--30 November 2024, as summarized in Table~\ref{tab:psr1937_obsid}.
In the GV closed condition in orbit, the exposure was extended to match the photon statistics expected in the open GV plan (80~ks). 
The expected $P$ and the period derivative $\dot{P}$ are $P=0.00155780656918537300$~s and $\dot{P}=-1.051003194988945 \times 10^{-19}$~s~s$^{-1}$ according to the results of previous X-ray observations with NICER (Neutron star Interior Composition ExploreR).\cite{2023AcAau.210..141S}
For comparison, we also took off-source data (OBSID = 000122000) with an exposure of about 60~ks before the observation of PSR~B1937+21.

\begin{table*}[hbt]
    \centering
    \caption{Summary of commissioning observations for timing verification.}
    \begin{tabular}{llcclc}
    \hline 
    OBSID & Object name & RA & Dec & Start time & Exp.$^\dagger$\\
    \hline 
    000122000 & PSR\_B1937+21\_OFFSET	 & 294.914 & 21.62205 & 2023-11-24 03:01:04  & 58.1 \\
    000123000 & PSR\_B1937+21 & 294.90959 & 21.5827 & 2023-11-25 08:31:04  & 245.1	 \\
    \hline 
    \end{tabular}
    \\
    {\scriptsize $\dagger$ Exposure in kiloseconds}
    \label{tab:psr1937_obsid}
\end{table*}

\subsection{Results}
\label{sec:timing_commissioning_results}  

\begin{figure}[htb]
    \centering
    \includegraphics[width=0.5 \textwidth]{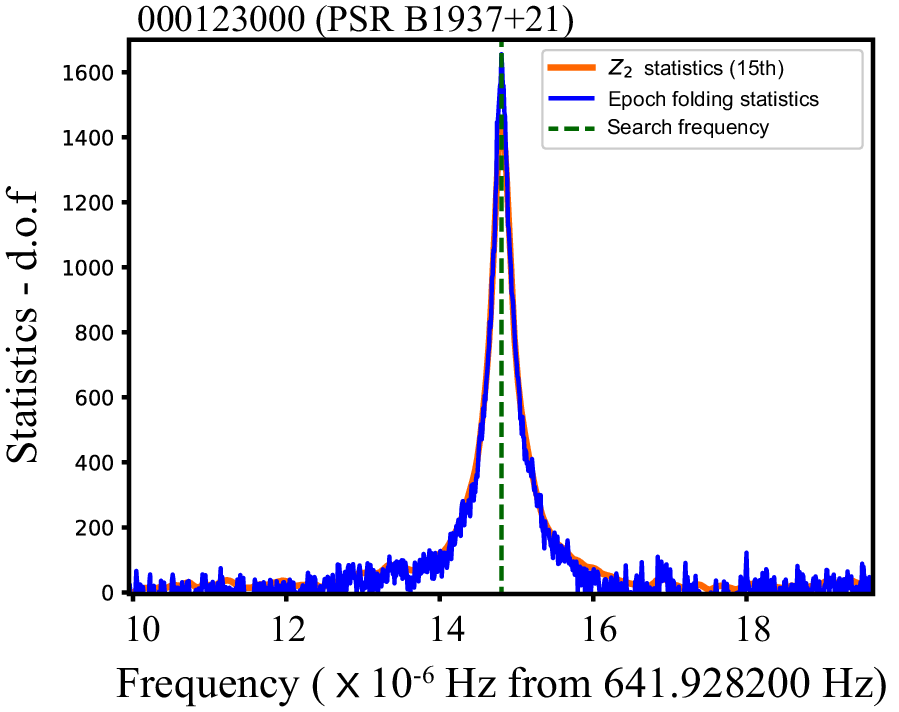}
    \caption{Periodogram for XRISM Resolve Hp events of PSR~B1937+21 with $Z_2$ statistics (red) and epoch-folding statistics (blue) in searches around frequency expected from the NICER ephemeris at the XRISM observation date} (green dashed line).
    \label{fig:commissioning_result_periodogram}
\end{figure}

We obtained about $2\times10^3$ and $4\times10^2$ events with Resolve in Hp grade from the object (OBSID = 000123000) and offset observation (OBSID = 000122000), respectively.
First, we applied barycentric correction at the target position (RA,~Dec) = (294.910672$^\circ$, 21.583090$^\circ$) using the HEAsoft tool ``barycen''\cite{2018JATIS...4a1206T} with the orbital elements determined for these observations.
Next, we used version~2.1 of the timing tool ``Stingray''\cite{2019ApJ...881...39H} to search for a possible periodicity in both data sets around the NICER period $P$ with fixed $\dot{P}$ described in Sec.~\ref{sec:timing_commissioning_observation}.
Consequently, a strong periodic signal was found only from the observation of PSR\_B1937+21 (OBSID = 000123000) in both the epoch-folding and $Z^2$ methods,\cite{2016ApJ...822...14B} as shown by the periodograms in Fig.~\ref{fig:commissioning_result_periodogram}.
The peak is detected with an accuracy of about $3\times 10^{-7}$ Hz at the frequency corresponding to the XRISM observation date (at 14.8 on Fig.~\ref{fig:commissioning_result_periodogram}), as extrapolated from the NICER ephemeris. 
Consequently, XRISM clearly observed the evolution of $P$ by $\dot{P} \sim 10^{-19}$~s~s$^{-1}$ with approximately a five-year interval after the NICER date (whose frequency corresponds to 978.8 on Fig.~\ref{fig:commissioning_result_periodogram}).
In the $Z^2$ statistical method, the periodicity becomes significant when higher harmonics up to order~15 are included, suggesting a very sharp pulse profile.

\begin{figure}[htb]
    \centering
    \includegraphics[width=0.5 \textwidth]{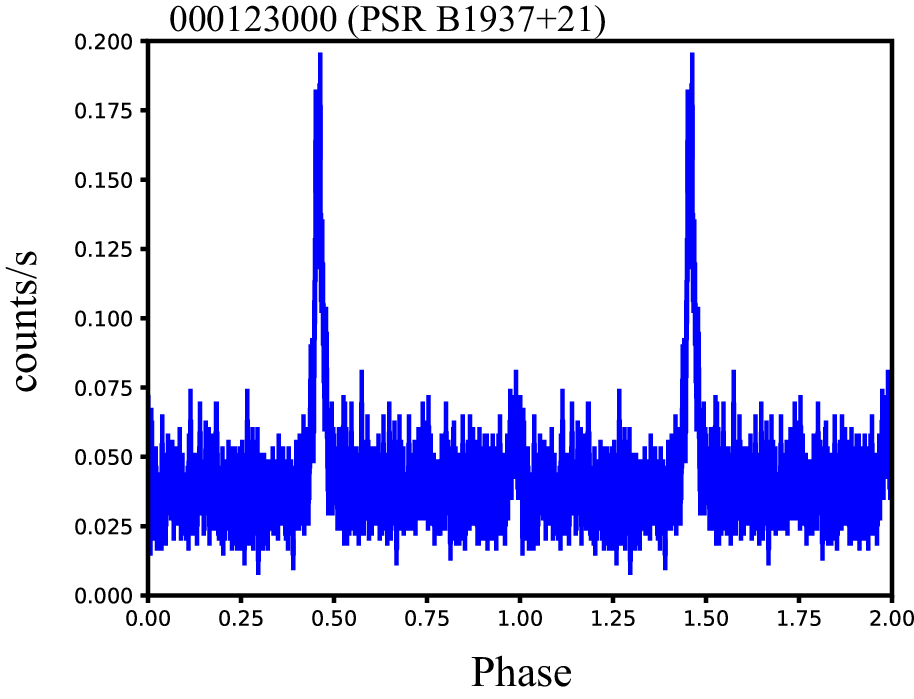}
    \caption{Pulse profile of PSR~B1937+21 using XRISM Resolve Hp events at NICER period $P$.}
    \label{fig:commissioning_result_efold}
\end{figure}

\begin{figure}[htb]
    \centering
    \includegraphics[width=0.50 \textwidth]{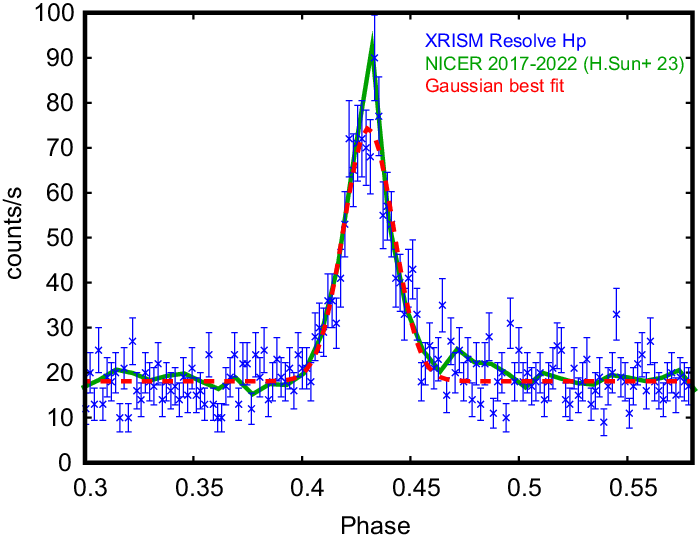}
    \caption{Same plot as Fig.~\ref{fig:commissioning_result_efold} right with XRISM Resolve Hp grade events folded by NICER period $P$. The pulse profile of NICER\cite{2023AcAau.210..141S} and the best-fit Gaussian plus constant model are shown by the green line and the red dashed line, respectively. The NICER profile in green is scaled to the XRISM profile for normalization.}
    \label{fig:commissioning_result2}
\end{figure}

We then checked the folded light curves in the NICER period $P$ using both data sets, and the results are shown in Fig.~\ref{fig:commissioning_result_efold}.
As can be seen, only the observation of PSR\_B1937+21 (OBSID = 000123000) results in a very sharp pulse profile at a phase of about 0.5. 
In comparison to the pulse profile with NICER, which has quite high timing accuracy in jitters at 100~ns in $1\sigma$, no significant degradation is observed in the XRISM profile, as shown by Fig.~\ref{fig:commissioning_result2}. 
Quantitatively, when describing the profile with a Gaussian plus constant component between phases 0.30 and 0.58, the pulse widths for XRISM and NICER are approximately $53.1 \pm 2.4~\mu$s and $45.4 \pm 0.9~\mu$s (errors represented in 1$\sigma$), respectively. They are consistent within the $3\sigma$ confidence level.
This estimation with a Gaussian function implies that the deterioration in the pulse widths from NICER to XRISM is about $7.7 \pm 2.5~\mu$s.
We further evaluated the degradation of the XRISM pulse profile using the NICER profile directly instead of the single Gaussian function; here, we generated the profiles smoothed by the Gaussian function with the various widths using the NICER profile and compared them with the XRISM profile to obtain the degradation on the pulse widths. 
The degradation with this method is quantified as $15.0 \pm 0.8~\mu$s.
This value seems to be reasonable compared with the Hitomi performance on the jitter of timing accuracy shown in Table~\ref{tab:error_budget} and Terada \etal (2018)\cite{2018JATIS...4a1206T}.

In this observation of PSR\_B1937+21, the number of pulsed Hp events is only 700 among a total of $2\times10^4$ events including lower-grade and off-pulse events.
Our results show the effectiveness of XRISM's timing capability, facilitating a high-sensitivity search for coherent periodic signals, particularly when the pulse profile is as sharp as the millisecond pulsar PSR\_B1937+21.

\section{Measurement of Absolute Timing Offset in PV and Calibration Period}
\label{sec:timing_crab}  

\subsection{Purpose of In-orbit Absolute Timing Calibration}
\label{sec:timing_crab_obs_purpose}  

Timing accuracy can be characterized by two main aspects: 1) the error in the timing offset from the absolute time, and 2) the jitter associated with the time tags.
Regarding these two elements, the calibration items addressing timing performance for XRISM include the following: 
1a) the delay in timing through the spacecraft bus system, particularly between GPSR and SMU, as well as between SMU and SpaceWire nodes, and the estimation of the systematic errors on this item;
1b) the timing jitter within the timing system of spacecraft bus part;
2a) the overall offset between the time of photon arrival and trigger time in the Resolve events impacted by the detector pulse processing, and the systematic errors on this item; 
2b) the timing jitter in the Resolve events, primarily due to residual errors in the relative timing calibration between event properties such as pixels, grades, and incident X-ray energies.
Item~1b) was examined in Sec.~\ref{sec:timing_commissioning_results} during the commissioning period, although the value in Sec.~\ref{sec:timing_commissioning_results} partly includes the errors of item~1a).
The details of the Resolve part in 2a) and 2b) are described in Sawada et al. (2025)\cite{2025JATIS_Sawada_Time}, which conclude that the these items fall within the error budget.
In this section we validate and quantify the performance of item~1a) for the bus and the ground component of the XRISM timing system.

We performed the timing measurements in the PV and calibration period in the nominal operation phase, as in step~3 defined in Sec.~\ref{sec:timing_system_develop}.
The main objectives of the timing measurement during the PV and calibration period were (i) to measure the absolute timing offset of the Resolve events with the final timing parameters measured on the ground, 
(ii) to verify the relative timing of Resolve events such as differences between the pixels and grades, (iii) to refine Resolve timing calibration parameters to improve the timing accuracy, and (iv) to finally conclude the fulfillment of the 1 ms requirement based on a comprehensive assessment of the residual timing errors. 
This paper, we concentrate on the objective (i), while Sawada et al. (2025)\cite{2025JATIS_Sawada_Time} discusses the objectives (ii), (iii), and (iv).
In a similar manner, the timing calibration for XRISM Xtend was conducted during the PV and calibration period using the same observations. 
The mission requirement for Xtend is less stringent compared to Resolve, with a detailed procedure to be outlined in a forthcoming paper. 
Consequently, the focus herein is on the timing offset measurement of the Resolve instrument.

\subsection{Observation Summary of Crab Simultaneous Observation Campaign}
\label{sec:timing_crab_obs}  

The planning of the in-orbit timing calibration target and operations was completed well before launch, as an activity of the In-Flight Calibration Planning (IFCP) group under the XRISM Collaboration.\cite{2020SPIE11444E..26M,2025JATIS_Eric_Cal}
The candidates for the in-orbit timing calibration are defined in Table~\ref{tab:commissioning_objects}, which was also used for the commissioning plan as described in Sec.~\ref{sec:timing_commissioning_observation}.
Based on visibility and priority, the Crab Pulsar was selected for in-orbit timing calibration.

\begin{table*}[hbt]
    \centering
    \caption{Summary of Crab simultaneous observations with XRISM, NICER, and NuSTAR in March 2024.}
    \begin{tabular}{llccllc}
    \hline 
    OBSID & Object name & RA & Dec & Start time & Exp.$^\dagger$\\
    \hline 
    \multicolumn{6}{l}{\bf XRISM}\\
    100006050 & Crab\_SWoffset & 83.6203 & 22.00251 & 2024-03-20 04:39:04  & 13.102 \\
    100006040 & Crab\_SEoffset & 83.64819 & 22.00182 & 2024-03-19 18:11:04 & 13.289 \\
    100006030 & Crab\_NWoffset & 83.62066 & 22.02802 & 2024-03-19 11:46:04  & 7.989 \\
    100006020 & Crab\_NEoffset\_1 & 83.6479 & 22.02734  & 2024-03-19 03:46:04  & 12.031 \\
    100006010 & Crab\_NEoffset\_2 & 83.66076 & 22.04336  & 2024-03-18 13:31:04  & 19.813 \\
    \hline 
    \multicolumn{6}{l}{\bf NICER}\\
    7013010101 & PSR\_B0531+21 & 83.632625 & 22.015167  & 2024-03-1903:41:09  & 1.228 \\
\hline 
    \multicolumn{6}{l}{\bf NuSTAR}\\
    11002303004 & Crab & 83.602500 & 21.971667 & 2024-03-18 23:01:09  & 4.983 \\
    \hline 
    \end{tabular}
    \\
    {\scriptsize $\dagger$ Exposure in kiloseconds}
    \label{tab:crab_obsid}
\end{table*}

Because the main pulses of Crab in the X-ray band arrive a few hundred microseconds earlier than the radio pulses\cite{2010ApJ...708..403M,HXD_Time} and the arrival time depends on the observation epoch due to the period drift by $\dot{P}$, we require simultaneous X-ray observations of the Crab Pulsar.
The X-ray instruments NICER and NuSTAR (Nuclear Spectroscopic Telescope Array) have good timing capabilities at $<400$~ns and about $100~\mu$s in the absolute timing offsets with 100~ns and $65~\mu$s jitters ($1\sigma$), respectively, according to private communication from the timing working group of the International Astronomical Consortium for High-Energy Calibration (IACHEC).\footnote{https://wikis.mit.edu/confluence/display/iachec/Timing} 
Therefore, we performed simultaneous observations of Crab using XRISM, NICER, and NuSTAR on 18--19 March 2024. 
The details of the observation campaign are summarized in Table~\ref{tab:crab_obsid}.
XRISM observed Crab several times with multiple pointing positions, and NICER and NuSTAR observations were performed in the same epoch. 

\subsection{Data Analyses of Crab Observations}
\label{sec:timing_crab_analyses}  

To derive the pulse profile of the Crab Pulsar, we used the Jodrell Band radio ephemeris\footnote{https://www.jb.man.ac.uk/pulsar/crab.html} reported for 15 March 2024.
The parameters are summarized in Table~\ref{tab:crab_db_ephemeris}.

\begin{table}[htb]
    \centering
    \caption{Jodrell Band radio ephemeris of Crab Pulsar on 15 March 2024.}
    \begin{tabular}{cl}
    \hline 
       $P$  &  0.033826063215951 s \\
       $\dot{P}$  & $4.195894883145 \times 10^{13}$ s s$^{-1}$ \\
       epoch & MJD 60384.00000032858952\\
       Solar system ephemeris & FK5 (DE-200) \\
       Barycentric position & (RA,~DEC) = (83.633218, +22.014464)\\
    \hline 
    \end{tabular}
    \label{tab:crab_db_ephemeris}
\end{table}

In the XRISM analyses, TIME was calculated using the final timing parameters determined on the ground.  
Resolve events with Hp grade were selected.
Throughout the observation, the radioisotope $^{55}$Fe on the filter wheel was constantly exposed for gain calibration purposes. We did not exclude the $^{55}$Fe line band from our analyses as this emission has an insignificant effect on the pulse profile.
Barycentric correction at the solar ephemeris DE-200 was then applied to the screened events using ``barycen'' in FTOOLS, as done in Sec.~\ref{sec:timing_commissioning_results}.
We then plotted the folded light curve with the radio ephemeris in Table~\ref{tab:crab_db_ephemeris} using the same timing tool (stingray version~2.1) as in Sec.~\ref{sec:timing_commissioning_results}.

In the NICER analyses, the unusual optical light leak during the daytime periods made it impossible to use the cleaned event FITS files within the standard pipeline products.
Therefore, we used the unscreened event files ("ufa"-type events) for our timing analyses, as the time stamps are accurate despite the unreliable energy scale, as per a private communication from NICER instrument members.
We applied barycentric correction using ``barycorr'' in FTOOLS to the ``ufa'' event files under the DE-200 JPL ephemeris with coordinate expressed in the FK5 reference frame to match the Jodrell Bank ephemeris, and we generated the folded light curve with the radio ephemeris in Table~\ref{tab:crab_db_ephemeris} using stingray version~2.1.

For NuSTAR observation, the latest clock correction table was applied in the calculation of TIME, and barycentric correction was performed with ``barycorr'' under the DE-200 JPL ephemeris coordinate.
Dead-time correction was then applied to the folded light curve of the standard-cleaned event files calculated by the stingray version~2.1.
The dead-time-corrected folded light curve was provided by a member of the NuSTAR instrument team.

\subsection{Results}
\label{sec:timing_crab_result}  

\begin{figure}[htb]
    \centering
    \includegraphics[width=0.9 \textwidth]{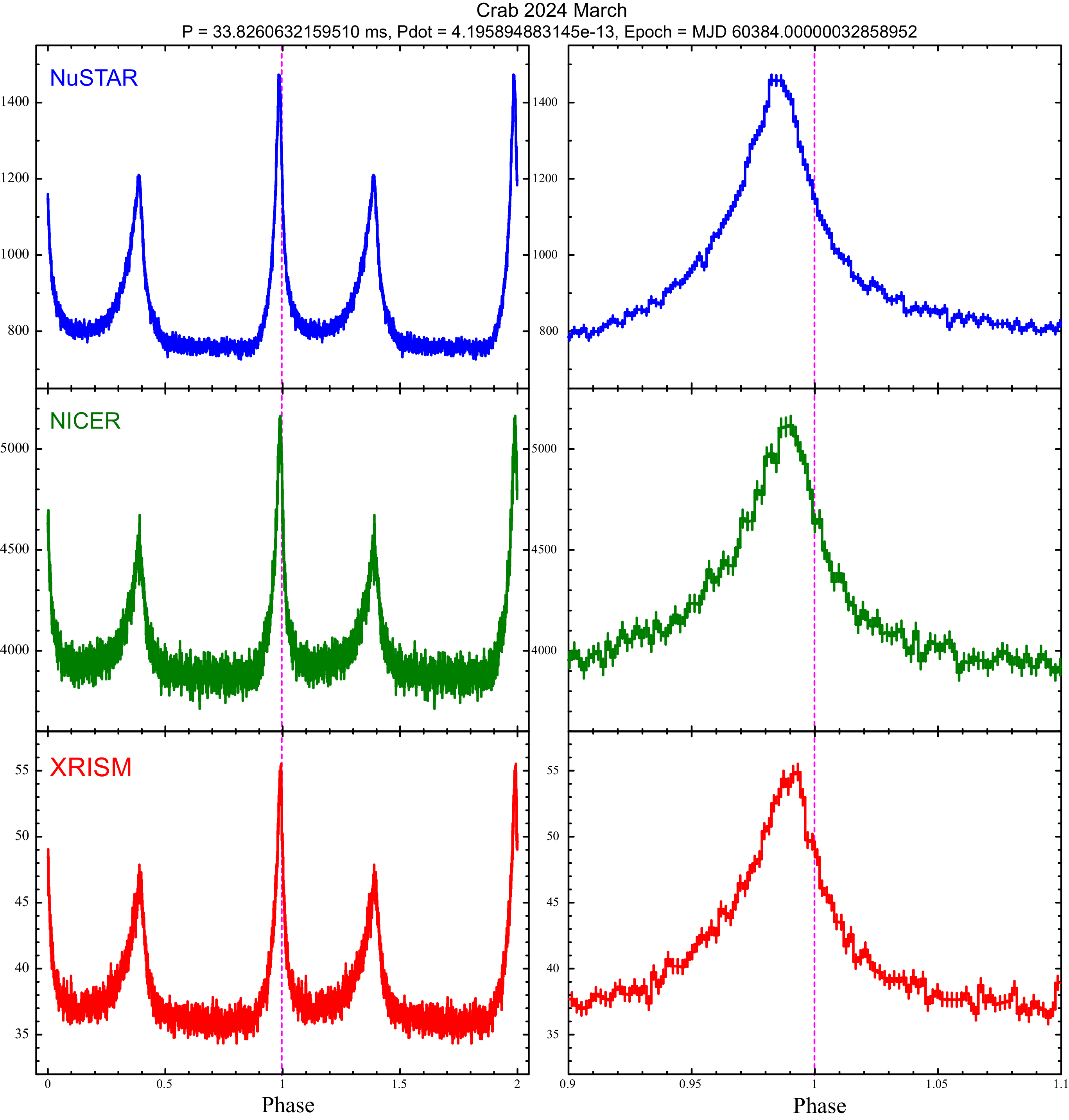}
    \caption{Pulse profiles of XRISM Resolve Hp (bottom row), NICER (middle row), and NuSTAR (top row) folded at Jodrell Bank radio ephemeris on 15 March 2024 in Table~\ref{tab:crab_db_ephemeris}. 
    The panels on the right show an expanded view around phase 1.0, which corresponds to the arrival time of the main pulse in the radio band.
    The vertical error bars on the data denote 1$\sigma$.}
    \label{fig:crab_result_pulseprofile}
\end{figure}

\begin{figure}[htb]
    \centering
    \includegraphics[width=0.9 \textwidth]{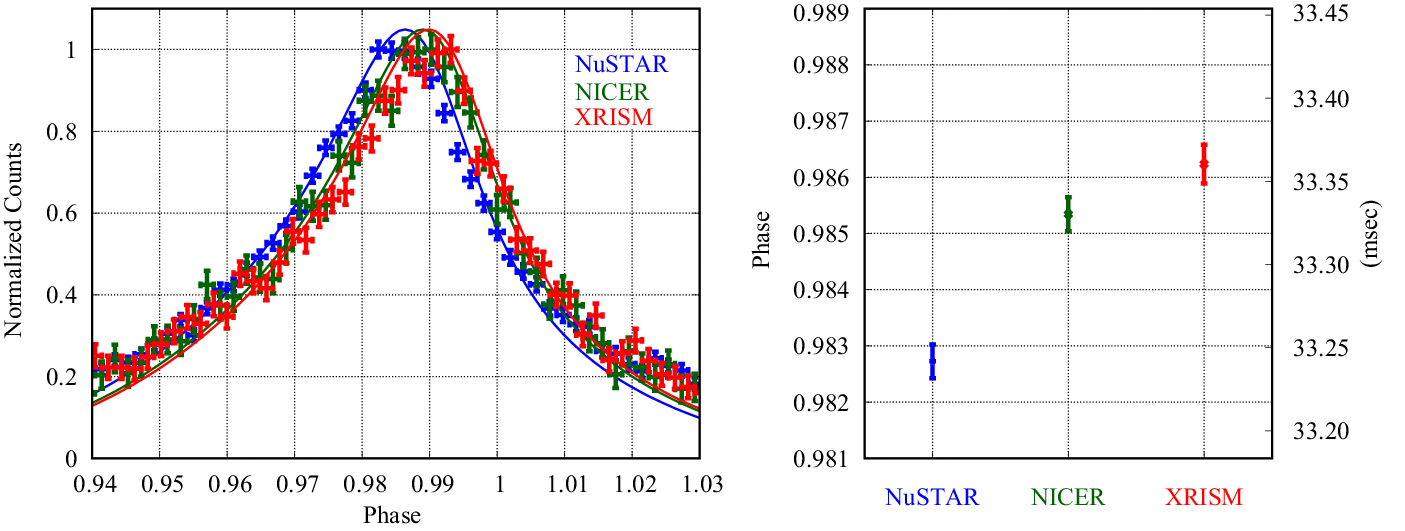}
    \caption{Left: same pulse profiles as in Fig.~\ref{fig:crab_result_pulseprofile} but with best-fit models described by Eq.~(1) of Ge \etal (2016)\cite{2016ApJ...818...48G}. 
    Right: offset phases of main pulses of XRISM, NICER, and NuSTAR. Errors represent in 1$\sigma$.}
    \label{fig:crab_result_offset}
\end{figure}

Figure~\ref{fig:crab_result_pulseprofile} shows the pulse profiles of Crab taken from the observation campaign with XRISM, NICER, and NuSTAR in March 2024. 
As represented in the panels on the right, the main pulse in the X-ray band comes a few hundred microseconds before the radio ephemeris (i.e., the phase $\phi = 0$).
The arrival time of the main pulse with XRISM (with timing parameters measured on the ground) is aligned with that of NICER or NuSTAR.
To test the difference in arrival time numerically, we normalized the pulse profiles and fitted in $0.93 < \phi < 1.03$ with Eq.~(1) of Ge \etal (2016)\cite{2016ApJ...818...48G} (originally from Nelson \etal 1970\cite{1970ApJ...161L.235N}), with the pulse parameters set to the RXTE PCA values listed in Table 2 of Ge \etal (2016)\cite{2016ApJ...818...48G}, and derived the phase of the main pulse [$\phi_0$ in Eq.~(1) of Ge \etal (2016)\cite{2016ApJ...818...48G}].
Figure~\ref{fig:crab_result_offset} (left) shows that the function reproduces the profiles well, and so the pulse parameters were derived by fitting. 
The phases $\phi_0$ from the radio ephemeris are summarized in Fig.~\ref{fig:crab_result_offset} (right) and indicate a difference of around $200~\mu$s between missions. The XRISM phase is aligned with the NICER phase within $100~\mu$s.

Consequently, the absolute timing offset of XRISM using the ground timing calibration, measured as the offset from the reference (NICER and NuSTAR) was confirmed to be consistent within $200~\mu$s; hence, the timing error originating from this component is determined to be $200~\mu$s.
Together with the timing jitter of Resolve Hp events evaluated at $15~\mu$s in Sec.~\ref{sec:timing_commissioning_results}, the timing accuracy of the combined bus and ground component in the XRISM timing system meets the timing error budget of $500~\mu$s.

\section{Conclusion}
\label{sec:conclusion}  

Including both on-board instruments and off-line data processing tools, the XRISM timing system has been designed to meet the scientific requirement described in Sec.~\ref{sec:timing_req}. 
Following the detailed design of the timing system as reported in Sec.~\ref{sec:timing_system}, we performed a timing verification experiment on the ground as reported in Sec.~\ref{sec:ground_test} and concluded that the timing performance in the nominal operating mode (GPS ON mode) satisfies the error budget for the bus and ground component of the XRISM timing system, i.e., items~A, B, and C in Table~\ref{tab:error_budget}.
Combined with the results in the GPS unsynchronized mode (GPS OFF mode) reported by Shidatsu \etal (2024,~2025)\cite{2024SPIE_Shidatsu,2025JATIS_Shidatsu_Time}, the bus and ground component of the timing system meets the error budget across all modes.
After the launch, we verified the timing capability in the full configuration by using the millisecond pulsar PSR~B1937+21 during the commissioning period in the initial operation phase as reported in Sec.~\ref{sec:timing_commissioning}, demonstrating that XRISM can detect the sharp periodic signal from such a rotating object.
By comparing the XRISM pulse profile of PSR~B1937+21 taken with that taken by NICER, we showed that the timing jitter of the bus and ground component of the XRISM timing system is below $15.0 \pm 0.8~\mu$s.
Then, in the PV and calibration period in the nominal operation phase, we performed simultaneous observations of the Crab Pulsar with NuSTAR and NICER as reported in Sec.~\ref{sec:timing_crab}, showing that the arrival time of the main pulse with XRISM is aligned with that with NICER and NuSTAR to within $200~\mu$s.
Together with the timing jitter (ca.  $15~\mu$s with PSR~B1937+21) and offset (ca.  $200~\mu$s with Crab) obtained in orbit, we concluded that the timing accuracy of the combined bus and ground component in the XRISM timing system meets the timing error budget of $500~\mu$s.
This is the conclusion of this article.

We found that the timing errors in the bus and ground components, mostly in jitter, are insignificant compared to the Resolve part, which is supposed to dominate the observed offset. Therefore, to conclude the achievement of the mission requirement on the 1 ms absolute timing accuracy, it is essential to have a comprehensive timing error assessment of the Resolve part, which is presented separately in Sawada et al. (2025) \cite{2025JATIS_Sawada_Time}.
Combining the paper series of the XRISM timing in this volume (this article, Shidatsu \etal\cite{2025JATIS_Shidatsu_Time}, and Sawada \etal\cite{2025JATIS_Sawada_Time}), the XRISM timing system satisfies the mission requirement.

\subsection*{Disclosures}
The authors declare that there are no financial interests, commercial affiliations, or other potential conflicts of interest that could have influenced the objectivity of this research or the writing of this paper.

\subsection* {Code, Data, and Materials Availability} 
The data presented in this article (ASCII data for Figures \ref{fig:ground_result}, \ref{fig:commissioning_result_periodogram}, \ref{fig:commissioning_result_efold}, and \ref{fig:crab_result_pulseprofile}) are publicly available in the Data ARchives and Transmission System (DARTS) Archive system at JAXA, \\
https://darts.isas.jaxa.jp/pub/xrism/papers/Terada2025/.

\subsection* {Acknowledgments}
The authors thank all the XRISM team members, particularly for their continuous efforts on spacecraft operation, maintenance of onboard instruments, and science operations, and thank the International Astronomical Consortium for High-Energy Calibration (IACHEC) for the coordination of the simultaneous observations of Crab using multiple X-ray missions in Section \ref{sec:timing_crab}. 
The authors thank Dr. Matteo Bachetti for providing the dead-time corrected pulse profile of the Crab with NuSTAR in Section \ref{sec:timing_crab}, and thank Dr. Craig B Markwardt for supporting the analyses of Crab observations with NICER in the same section.
This work was supported by the JSPS Core-to-Core Program (grant number: JPJSCCA20220002) and Japan Society for the Promotion of Science Grants-in-Aid for Scientific Research (KAKENHI) Grant Number JP20K04009 (YT), JP19K14762, JP23K03459, JP24H01812 (MS), JP24K17105 (YK), and JP21K20372 (TY). 
The material is based on work supported by NASA under award numbers 80GSFC21M0002 (KP, ML), 80NSSC20K0737 (EDM), and 80NSSC24K0678 (EDM) and by the Strategic Research Center of Saitama University (MST,YT,SK).
The authors thank the English proofreading service, ThinkSCIENCE, and an AI tool, Writeful, for improving the language quality of this paper.
The authors appreciate the insightful comments and suggestions provided by the anonymous reviewers and the editors, M. Ozaki \etal.

\vspace{2ex}\noindent\textbf{Yukikatsu Terada} is a professor at Saitama University and the Japan Aerospace Exploration Agency (JAXA). He received his BS and MS degrees in physics, and a PhD degree in science from the University of Tokyo in 1997, 1999, and 2002, respectively. He is the leader of the Science Operations Team of the X-Ray Imaging and Spectroscopy Mission (XRISM). He is chairing the timing working group in the International Astronomical Consortium for High-Energy Calibration (IACHEC).

\vspace{2ex}\noindent\textbf{Megumi Shidatsu} is an associate professor at Ehime University. She received her BS, MS, PhD degrees in science from Kyoto University in 2010, 2012, and 2015, respectively. She is a member of the Science Operations team of the X-Ray Imaging and Spectroscopy Mission (XRISM) project. She is also a member of the timing working group in the International Astronomical Consortium for High-Energy Calibration (IACHEC) and is involved in cross-timing calibration among X-ray missions.

\vspace{2ex}\noindent\textbf{Makoto Sawada} is an assistant professor at Rikkyo University. He received his BS, MS, and PhD degrees in science from Kyoto University in 2007, 2009, and 2012, respectively. He is a member of the Resolve team and the leader of the calibration implementation team of the X-Ray Imaging and Spectroscopy Mission (XRISM) project. His current research interests include spectroscopy of high-energy astrophysical plasmas and instrumentation for high-resolution X-ray spectroscopy.

\vspace{1ex}
\noindent Biographies and photographs of the other authors are not available.

\listoffigures
\listoftables

\end{spacing}
\end{document}